\documentclass[aps,prc,twocolumn,superscriptaddress,showpacs]{revtex4-1}

\usepackage{graphicx,amsmath,amssymb,bm}
\usepackage[T1]{fontenc}
\newcommand{\mev}{\,\mathrm{MeV}}

\newcommand{\fmiq}{\,\mathrm{fm}^{-3}}
\newcommand{\kf}{k_\mathrm{F}}

\begin{document}

\title{The chiral condensate in neutron matter}
\author{T.\ Kr\"uger}
\email[E-mail:~]{thomas.krueger@physik.tu-darmstadt.de}
\address{Institut f\"ur Kernphysik, Technische Universit\"at Darmstadt,
64289 Darmstadt, Germany}
\affiliation{ExtreMe Matter Institute EMMI, GSI Helmholtzzentrum f\"ur
Schwerionenforschung GmbH, 64291 Darmstadt, Germany}
\author{I.\ Tews}
\email[E-mail:~]{tews@theorie.ikp.physik.tu-darmstadt.de}
\affiliation{Institut f\"ur Kernphysik, Technische Universit\"at Darmstadt,
64289 Darmstadt, Germany}
\affiliation{ExtreMe Matter Institute EMMI, GSI Helmholtzzentrum f\"ur
Schwerionenforschung GmbH, 64291 Darmstadt, Germany}
\author{B.\ Friman}
\email[E-mail:~]{b.friman@gsi.de}
\affiliation{GSI Helmholtzzentrum f\"ur Schwerionenforschung GmbH, 
64291 Darmstadt, Germany}
\author{K.\ Hebeler}
\email[E-mail:~]{kai.hebeler@physik.tu-darmstadt.de}
\affiliation{Institut f\"ur Kernphysik, Technische Universit\"at Darmstadt, 
64289 Darmstadt, Germany}
\affiliation{ExtreMe Matter Institute EMMI, GSI Helmholtzzentrum f\"ur 
Schwerionenforschung GmbH, 64291 Darmstadt, Germany}
\affiliation{Department of Physics, The Ohio State University, 
Columbus, OH 43210, USA}
\author{A.\ Schwenk}
\email[E-mail:~]{schwenk@physik.tu-darmstadt.de}
\affiliation{ExtreMe Matter Institute EMMI, GSI Helmholtzzentrum f\"ur
Schwerionenforschung GmbH, 64291 Darmstadt, Germany}
\affiliation{Institut f\"ur Kernphysik, Technische Universit\"at Darmstadt,
64289 Darmstadt, Germany}

\begin{abstract}
We calculate the chiral condensate in neutron matter at zero temperature
based on nuclear forces derived within chiral effective field theory.
Two-, three- and four-nucleon interactions are included consistently
to next-to-next-to-next-to-leading order (N$^3$LO) of the chiral
expansion. We find that the interaction contributions lead to a modest
increase of the condensate, thus impeding the restoration of chiral
symmetry in dense matter and making a chiral phase transition in
neutron-rich matter unlikely for densities that are not significantly
higher than nuclear saturation density.
\end{abstract}

\pacs{21.65.Cd, 12.39.Fe, 21.30.-x, 26.60.-c}

\maketitle

The understanding of the phase diagram of matter is a current frontier
in nuclear physics. At high temperatures and vanishing net baryon
density, the properties of strongly interacting matter have been
studied in first-principle lattice QCD calculations. It is found that
at a temperature of $154 \pm 9 \mev$ matter exhibits a chiral and
deconfinement crossover transition from the low-temperature hadronic phase,
where chiral symmetry is spontaneously broken, to the chirally
symmetric high-temperature phase, the quark-gluon plasma~\cite{Bazavov}.
An order
parameter for characterizing this transition is the chiral
condensate.~\cite{Drukarev,Cohen,Lutz}.

Owing to the fermion sign problem, matter at non-zero baryon densities
cannot be directly studied in lattice QCD. Therefore, there are no
first-principle QCD results for the phase diagram in particular at low
temperatures and high densities. These conditions are probed in
neutron stars, which can reach several times nuclear saturation
density in their interiors~\cite{Haensel,Hebeler,Lattimer,longns}.
However, there has been much speculation on possible exotic phases
that may appear in the center of neutron stars, such as Lee-Wick
abnormal matter, pion- as well as kaon-condensed matter, hyperon
matter and quark matter~\cite{Haensel}. It has also been conjectured, see, e.g., Fig. 1 in Ref.~\cite{Sagert}, 
that in neutron matter, the QCD phase transition may begin already 
below saturation density.

Recent observations of neutron stars with $2 \, M_\odot$
masses~\cite{Demorest,Antoniadis} provide general constraints on the
equation of state (EOS) of cold strongly interacting matter, and put
into question whether exotic phases that tend to soften the EOS are realized
in neutron stars. At densities $n \lesssim n_0$, where $n_0 = 0.16
\fmiq$ denotes nuclear saturation density, the properties of nuclear
systems have been studied systematically based on nuclear forces derived
within chiral effective field theory (EFT)~\cite{RMP,Hammer} and using
renormalization group methods~\cite{PPNP,RPP}. In this paper, we use 
chiral EFT interactions to study the chiral condensate as a function
of density in neutron matter, based on perturbative calculations around
the first-order Hartree-Fock energy~\cite{neutmatt,fullN3LO,longN3LO}.

The chiral condensate can be obtained from the energy using the
Hellman-Feynman theorem~\cite{Cohen,Kaiser,Weise},
\begin{equation}
\langle \bar{q}q \rangle_n - \langle \bar{q}q \rangle_0 = 
n \, \frac{\partial}{\partial m_q} \left[\frac{E_\text{free}
(m_{q}, \kf)}{N}+\frac{E_\text{int}(m_{q}, \kf)}{N} \right] ,
\label{eq:hellman-feynman}
\end{equation}
where $\langle\bar{q}q\rangle_n$ and $\langle\bar{q} q \rangle_{0}$
are the chiral condensates at finite baryon density $n =
\kf^3/(3 \pi^2)$ (with Fermi momentum $\kf$) and in vacuum,
respectively.  Moreover, $E_\text{free}/N = m_N + 3 \kf^2/(10m_N)$ is
the energy per particle of a system consisting of $N$ noninteracting
degenerate neutrons in the nonrelativistic limit, $E_\text{int}$ is
the corresponding interaction energy, $m_q$ denotes the average of the
$u$ and $d$ quark masses, $\bar{q} q = \bar{u} u + \bar{d} d$, and
$m_N$ is the neutron mass.

The contribution from the nucleon mass to the chiral condensate is
proportional to the pion-nucleon sigma term $\sigma_{\pi N}$, which
accounts for the scalar quark density in the nucleon~\cite{Cohen,Gasser}:
\begin{equation}
\sigma_{\pi N} = \left< N \right| m_q \bar{q}q \left| N \right> 
= m_q \frac{\partial m_N}{\partial m_q} \,.
\end{equation}
Here $\left| N \right>$ represents the state of a nucleon at rest. The
value of the pion-nucleon sigma term has been determined within
different frameworks~\cite{Gasser,Frink,Semke,Alarcon,Stahov}. As a baseline
we use the value $\sigma_{\pi N} \approx 50 \mev$~\cite{Weise}. The
chiral condensate in neutron matter relative to the vacuum is then
given by~\cite{Cohen}
\begin{align}
\frac{\langle\bar{q} q \rangle_{n}}{\langle\bar{q} q \rangle_{0}}
&= 1-\frac{n}{f_{\pi}^2}\frac{\sigma_{\pi N}}{m_{\pi}^2}
\left(1-\frac{3 \kf^2}{10 m_N^2} + \ldots \right) \nonumber \\[1mm]
&\quad -\frac{n}{f_{\pi}^2}\frac{\partial}{\partial m_{\pi}^2}
\frac{E_\text{int}(m_{\pi}, \kf)}{N} \,,
\label{eq:cc}
\end{align} 
where we have used the Gell-Mann--Oakes--Renner relation $m_q \langle
\bar{q}q\rangle_0 = -f_\pi^2m_\pi^2$. In our calculation we use for
the pion mass the charge average $m_{\pi} = 138 \mev$ and for the
pion decay constant $f_{\pi} = 92.4 \mev$.

\begin{figure}[t]
\begin{center}
\includegraphics[width=\columnwidth,clip=]{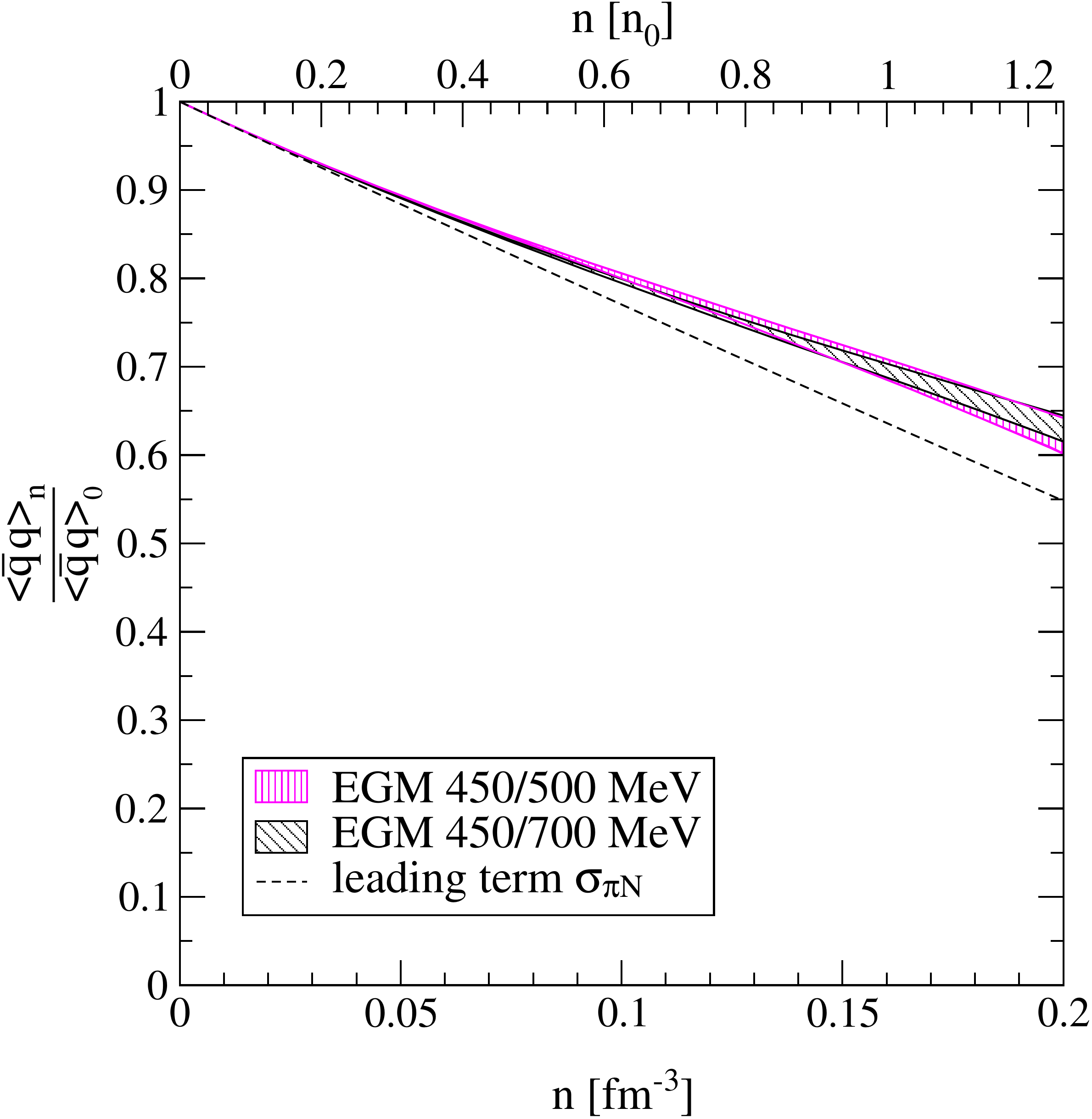}
\end{center}
\caption{(Color online) Chiral condensate $\langle \bar{q} q \rangle_{n}/
\langle \bar{q} q \rangle_{0}$ as a function of density in neutron matter.
The dashed line is the leading pion-nucleon sigma-term contribution. The
interaction contributions are obtained from the N$^3$LO neutron-matter
calculation of Refs.~\cite{fullN3LO,longN3LO}, based on the EGM $450/500
\mev$ and $450/700 \mev$ N$^3$LO NN potentials plus 3N and 4N
interactions to N$^3$LO, by varying the pion mass around the physical
value. As in Refs.~\cite{fullN3LO,longN3LO}, the bands for each NN
potential include uncertainties of the many-body calculation, of the
$c_i$ couplings of 3N forces, and those resulting from the 3N/4N cutoff
variation.\label{fig:total}}
\end{figure}

The leading $\sigma_{\pi N}$ contribution to the chiral condensate in
Eq.~(\ref{eq:cc}), which is due to the mass term in $E_\text{free}/N$,
is linear in density and is shown in Fig.~\ref{fig:total} by the
dashed line. By extrapolating this linear density dependence, one
finds restoration of chiral symmetry at a density around $(2.5-3)
n_0$~\cite{Cohen,Kaiser}. For the density range shown in Fig.~\ref{fig:total},
where chiral EFT interactions can be applied with confidence, the
kinetic energy contribution to $E_\text{free}/N$ gives only a $4 \%$
correction relative to the leading term. Relativistic
corrections, indicated by the dots in Eq.~(\ref{eq:cc}), are 
negligible at these densities.  The next term $9 k_F^4/56 m_N^4$ is a 0.3 \% correction. While the first correction to the
chiral condensate in Eq.~(\ref{eq:cc}) is a consequence of the finite
nucleon density, the long-range contributions from $E_{\rm int}$ can be
attributed to the modification of the scalar pion density $\Delta n_{\pi}^{s} = n \, \partial(
E_\text{int}/N)/\partial m_\pi$ due to the interactions between 
nucleons (cf. Ref.~\cite{Friman}).

The energy per particle of neutron matter has recently been calculated
based on chiral EFT interactions to N$^3$LO, including two-nucleon
(NN), three-nucleon (3N), and four-nucleon (4N)
forces~\cite{fullN3LO,longN3LO}. The pion-mass dependence of nuclear
forces arises from two sources: First, due to the explicit $m_\pi$
dependences in the long-range pion-exchange interactions, and second,
implicitly, due to the quark-mass dependence of the pion-nucleon
coupling $g_A$, the pion decay constant $f_\pi$, as well as the
leading NN contact interactions $C_S$ and $C_T$, and higher-order
pion-nucleon and short-range NN and 3N contact interactions.

We calculate the explicit $m_\pi$ dependence of nuclear forces by
varying the value of the pion mass in the pion-exchange NN, 3N, and 4N
interactions. At the NN level, we use the N$^3$LO potentials of
Epelbaum, Gl\"ockle, and Mei{\ss}ner (EGM)~\cite{EGM1,EGM2} with
cutoffs 450/500 and 450/700 MeV (and their N$^2$LO versions to study
the order-by-order convergence). With these NN interactions neutron
matter is perturbative at the densities considered
here~\cite{fullN3LO,longN3LO}. This result was recently validated 
using first Quantum Monte Carlo calculations with chiral EFT
interactions~\cite{QMC}. We vary $m_\pi$ by $0.5\%$ in the corresponding
potential routines. The derivative of the interaction energy with respect
to $m_\pi^2$ in Eq.~(\ref{eq:cc}) is then computed numerically for
different densities.

\begin{figure*}[t]
\hspace*{0.35cm}
\includegraphics[height=6.0cm,clip=]{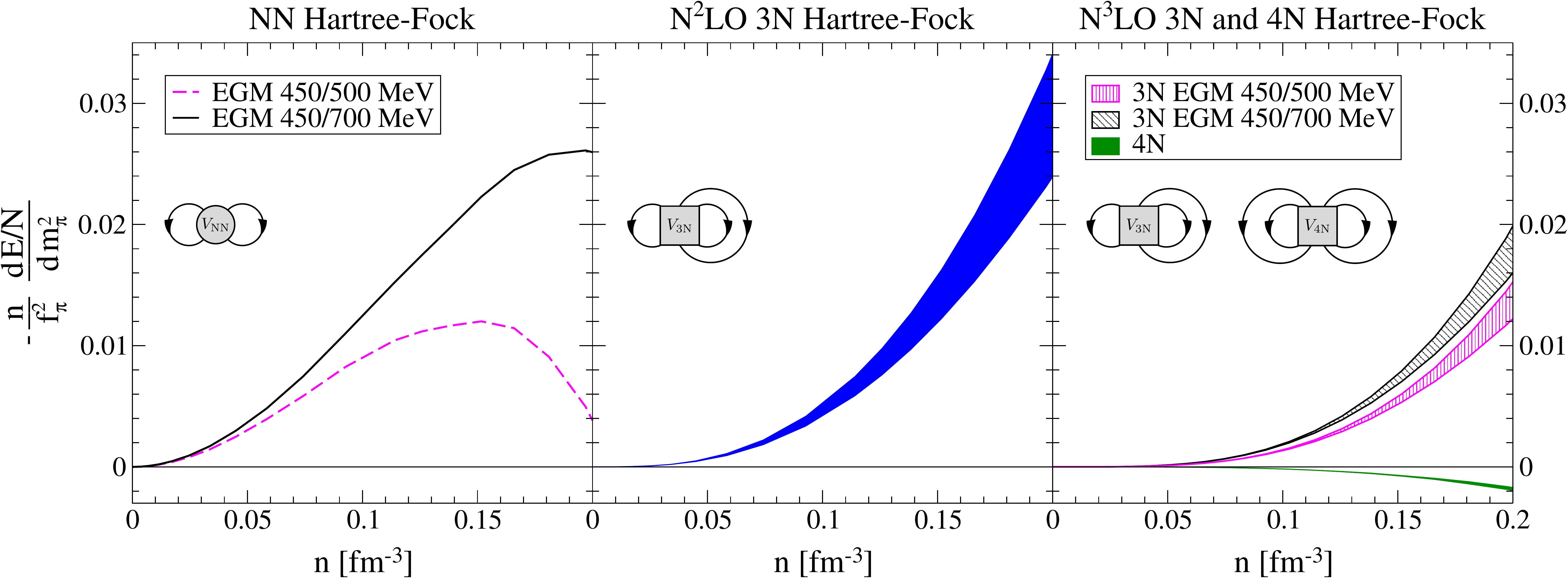}
\newline\newline
\includegraphics[height=6.12cm,clip=]{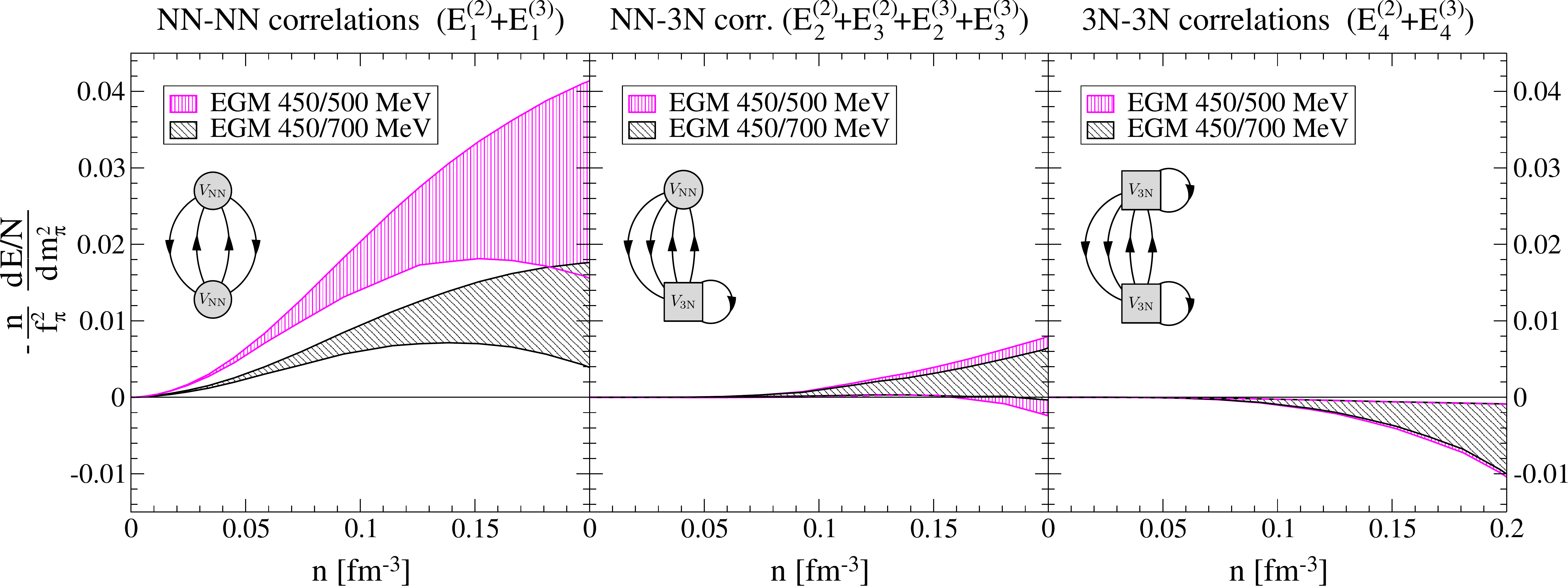}
\caption{(Color online) Individual interaction contributions to the
chiral condensate in neutron matter for the two N$^3$LO NN potentials
of Fig.~\ref{fig:total}. The upper row gives the NN, the N$^2$LO 3N,
and the N$^3$LO 3N and 4N Hartree-Fock contributions. In the lower
row, second-order and particle-particle/hole-hole third-order
contributions beyond Hartree-Fock are shown, where the N$^2$LO 3N forces
are included as density-dependent two-body interactions.  The
various contributions are illustrated diagrammatically and the 
$E_i^{(2,3)}$ nomenclature follows Ref.~\cite{longN3LO}. The 
Hartree-Fock 3N- and 4N-force contributions include uncertainty 
estimates from the 3N/4N cutoff variation and from the $c_i$ 
couplings of 3N forces. The higher-order bands also include 
uncertainties in the many-body
calculation.\label{fig:individual}}
\end{figure*}

\begin{figure*}[t]
\begin{center}
\includegraphics[height=6.1cm,clip=]{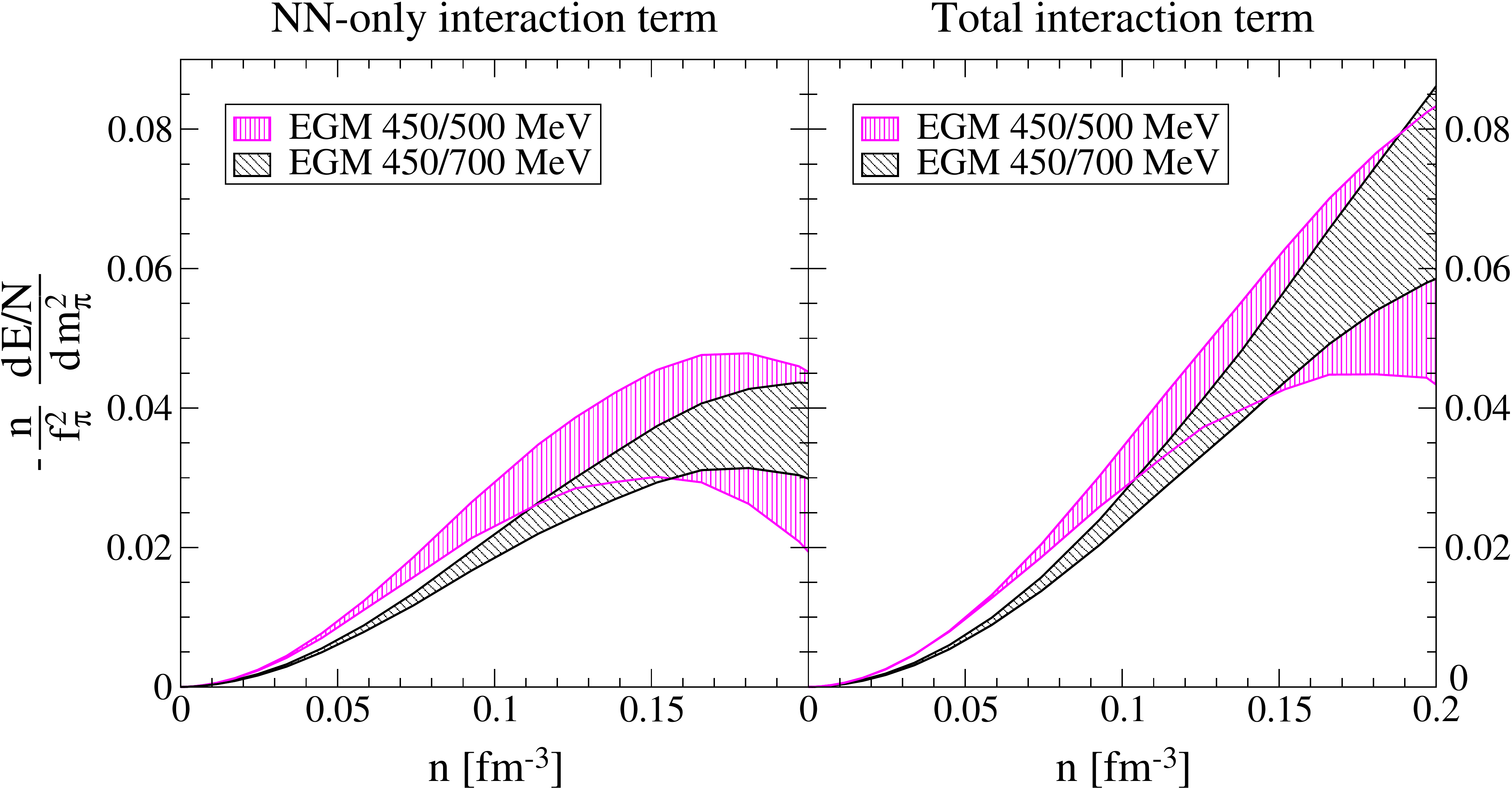}
\end{center}
\caption{(Color online) Sums of the NN-only and of the total interaction
contributions to the chiral condensate in neutron matter. The bands are
based on Fig.~\ref{fig:individual} and include the various uncertainty
estimates.\label{fig:sums}}
\end{figure*}

We estimate the impact of the quark-mass dependence of $g_A$ and
$f_\pi$ using the results of Refs.~\cite{Bernard,Berengut}. In a
perturbative calculation, the interaction energy per particle
$E_\text{int}/N$ is a polynomial in $g_A$ and in $1/f_\pi$. Consider a
term in this polynomial, in which $g_A$ enters with the power
$\alpha$, $\left[E_\text{int}/N\right]_{\alpha}$. For the
corresponding contribution to the chiral condensate, due to the
pion-mass dependence of $g_A$, we thus have
\begin{align}
&\quad -\frac{n}{f_\pi^2} \frac{\partial}{\partial m_\pi^2}
\left[\frac{E_\text{int}}{N}\right]_{\alpha}
=-\frac{n}{f_\pi^2}\frac{\partial g_A}{\partial m_\pi^2} 
\frac{\alpha}{g_A} \left[\frac{E_\text{int}}{N}\right]_{\alpha} 
\nonumber \\[1mm]
&\approx -(4.4-5.7) \times 10^{-4} \mev^{-1} \:
\alpha \left[\frac{E_\text{int}}{N}
\right]_{\alpha} \,.
\label{eq:nu_corr}
\end{align}
Here, we have used the range for $\frac{\partial g_A}{ \partial
m_\pi^2}$ from Ref.~\cite{Berengut}. Similarly, an interaction term,
in which $1/f_\pi$ enters with the power $\beta$, $\left[E_\text{int}/N
\right]_{\beta}$, leads to a contribution to the chiral condensate
\begin{align}
&\quad -\frac{n}{f_\pi^2}\frac{\partial}{\partial m_\pi^2}
\left[\frac{E_\text{int}}{N}\right]_{\beta}
=-\frac{n}{f_\pi^2}\frac{\partial f_\pi}{\partial m_\pi^2}
\frac{\partial}{\partial f_\pi}
\left[\frac{E_\text{int}}{N}\right]_{\beta} \nonumber \\[1mm]
&\approx (2.6-5.0) \times 10^{-4} \mev^{-1} \:
\beta \left[\frac{E_\text{int}}{N}\right]_{\beta} \,,
\label{eq:mu_corr}
\end{align}
using $\frac{\partial f_\pi}{ \partial m_\pi^2}$ from
Ref.~\cite{Berengut}. The uncertainty is larger in this case, because
of the $c_3$ and $c_4$ uncertainties, which are taken as in the
N$^3$LO calculations of Refs.~\cite{fullN3LO,longN3LO}.

For the leading-order one-pion-exchange NN interaction, which is
proportional to $g_A^2/f_\pi^2$ and contribute $\sim$$\,10 \mev$ per
particle at $n_0$, the terms~(\ref{eq:nu_corr}) and~(\ref{eq:mu_corr})
give a contribution to the chiral condensate ranging from $-0.006$ to
$+0.001$. The leading N$^2$LO two-pion-exchange 3N forces also provide
$\sim$$\,10 \mev$ per particle at $n_0$. These terms are proportional to
$g_A^2/f_\pi^4$ and the corresponding contribution to the chiral
condensate lies in the range $-0.001$ to $+0.011$. Combined, these 
corrections amount to at most a $25 \%$ increase of the uncertainty
band in Fig.~\ref{fig:total}. We expect the contributions from the
shorter-range interactions to start at a similar level. However, the
extrapolation of their $m_\pi$ dependence from lattice QCD results at
heavier pion masses to the physical point is uncertain. This will be 
improved in the future once lattice QCD results for NN and 3N interactions
for physical pion masses will become available. Because the estimated 
effects beyond the explicit $m_\pi$ dependence are small compared to the
band in Fig.~\ref{fig:total}, we do not include these contributions in
the present paper.

In Fig.~\ref{fig:total} we show our results for the chiral condensate
in neutron matter at N$^3$LO, based on the two N$^3$LO NN potentials
and including 3N and 4N interactions to N$^3$LO. The calculations
include all interactions at the Hartree-Fock level plus N$^3$LO NN and
N$^2$LO 3N interactions to second order and including
particle-particle/hole-hole third-order contributions, using a free or
a Hartree-Fock spectrum, as in Refs.~\cite{fullN3LO,longN3LO}. The bands
include the uncertainties of the many-body calculation, of the $c_i$
couplings of 3N forces, and those resulting from the 3N/4N cutoff
variation (see Refs.~\cite{fullN3LO,longN3LO} for details). The width 
of the bands are dominated by the uncertainties of the $c_3$ coupling
and by the sensitivity of the many-body calculation on the single-particle
spectrum used.

We find that the density dependence of the chiral condensate is
dominated by the leading $\sigma_{\pi N}$ term and therefore the
chiral condensate decreases almost linearly with increasing
density. The interaction contributions lead to a positive correction,
thus impeding the restoration of chiral symmetry with increasing
density. Consequently, for moderate densities, below say $n=0.3 \fmiq$,
which is below the linear extrapolation $n=(2.5-3) n_0$, a chiral
phase transition seems unlikely in neutron matter. However, we note
that, based on calculations of the type presented here, where only the
broken symmetry phase is considered, we cannot exclude the possibility
of a first-order transition, where the order parameter changes
discontinuously. If the transition occurs below $n=0.3 \fmiq$, this
would have to be a very strong first-order transition.

The results for the chiral condensate in neutron matter based on the
two EGM $450/500 \mev$ and $450/700 \mev$ N$^3$LO potentials are in
very good agreement within the uncertainty bands in Fig.~\ref{fig:total}.
At nuclear saturation density, $\langle\bar{q} q
\rangle_{n}/\langle\bar{q} q \rangle_{0}$ lies in the range $67.3-69.8
\%$ and $67.8-69.5 \%$, respectively. In comparison, the uncertainty
of the leading $\sigma_{\pi N}$ term is much larger. Using $\Delta
\sigma_{\pi N} = 8 \mev$~\cite{Gasser}, we find an uncertainty on the
order of $10 \%$ at $n=0.2\fmiq$.

The individual interaction contributions to the chiral condensate are
shown in Fig.~\ref{fig:individual}. In the upper row, the Hartree-Fock
N$^3$LO NN, N$^2$LO 3N, and N$^3$LO 3N and 4N results are given.  In
the lower row, second-order and particle-particle/hole-hole
third-order contributions beyond Hartree-Fock are shown, grouped into
the different contributions where N$^2$LO 3N forces are included as
density-dependent two-body interactions. This follows the notation of
Ref.~\cite{longN3LO}. The most important contributions are the NN and
3N Hartree-Fock terms as well as higher-order correlation effects due
to NN interactions. The latter are sensitive to the single-particle
spectrum used. This is because the Hartree-Fock single-particle
energies depend on the pion mass, so that the derivative with respect
to $m_\pi^2$ yields additional contributions. For the Hartree-Fock
spectrum, the NN correlation contributions are then only about half as
large as for the free spectrum. For the NN-3N and 3N-3N correlation
contributions, we find a similar sensitivity, but they are relatively small.

As shown in Fig.~\ref{fig:individual}, at nuclear saturation density
the NN Hartree-Fock contribution of the EGM 450/500 MeV potential is
smaller than that of the EGM 450/700 MeV potential by a factor
two. However, for the higher-order correlations the situation is
reversed, so that the sum of the Hartree-Fock and higher-order NN
contributions of the two NN potentials are in very good agreement, as
shown in the left panel of Fig.~\ref{fig:sums}. The total interaction
contribution, including 3N and 4N forces, is shown in the right panel
of Fig.~\ref{fig:sums} and yields a $6 \pm 2\%$ enhancement of the
chiral condensate at saturation density. We again find a very good
agreement within the uncertainty bands of the two NN potentials.

\begin{figure}[t]
\begin{center}
\includegraphics[width=0.85\columnwidth,clip=]{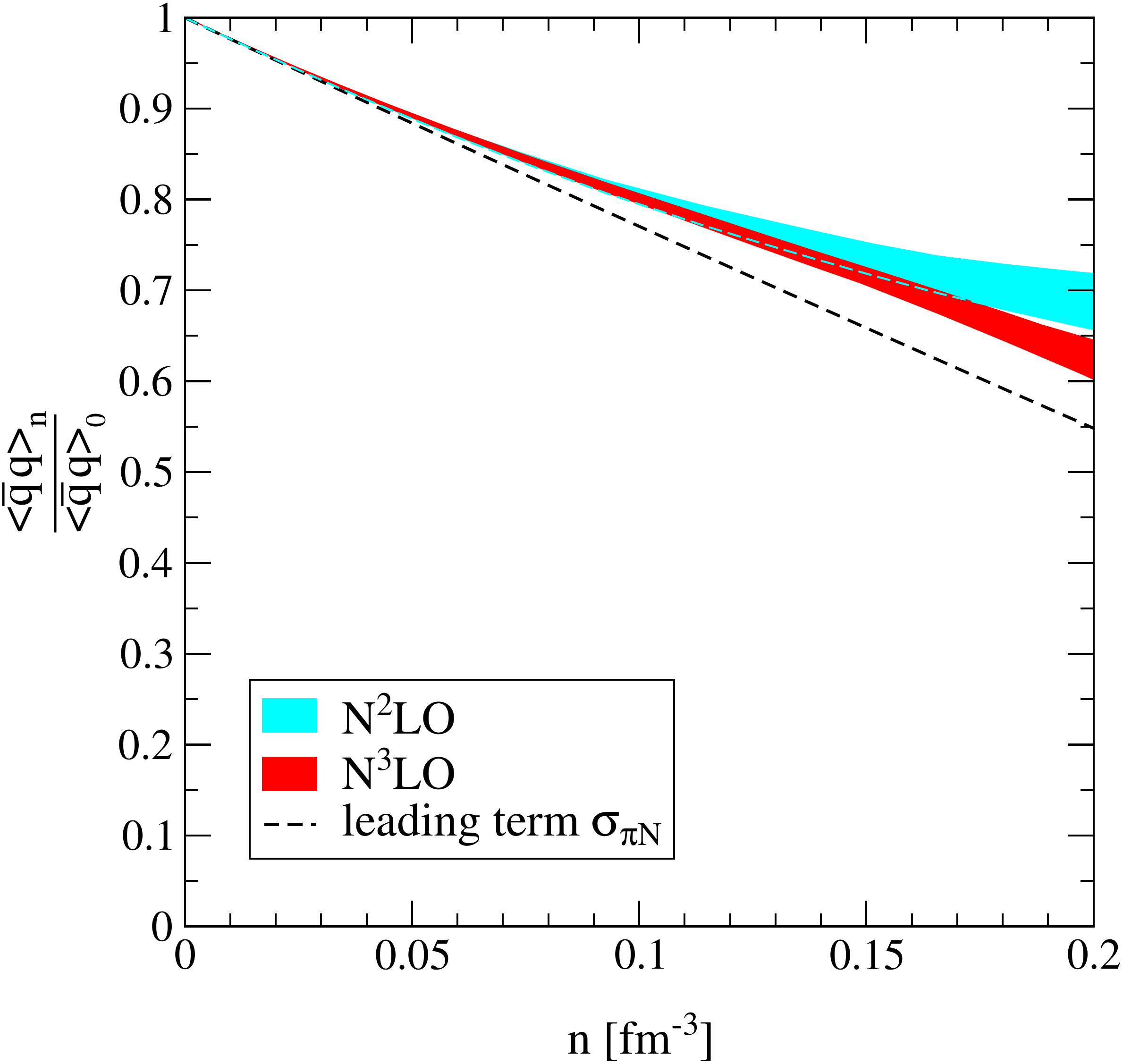}
\end{center}
\caption{(Color online) Chiral condensate as a function of density
in neutron matter at N$^2$LO and N$^3$LO in chiral EFT. The bands are
obtained as in Fig.~\ref{fig:total}.\label{fig:comparison}}
\end{figure}

The order-by-order convergence of the chiral EFT calculation for the
chiral condensate is shown in Fig.~\ref{fig:comparison} from N$^2$LO
to N$^3$LO. Going from N$^2$LO to N$^3$LO, the enhancement of the
condensate is weakly reduced and the width of the uncertainty band is
reduced by roughly a factor of two. A similar reduction of the bands
was found in the N$^3$LO calculation of the equation of
state~\cite{fullN3LO,longN3LO}.

We note that the increase of the chiral condensate due to interactions
corresponds to a decrease of the scalar pion density in the
interacting system. While the iterated NN one-pion-exchange
interaction yields an enhancement of the scalar pion
density~\cite{Friman}, the interference between the one-pion-exchange
and shorter-range NN parts induces the opposite effect. Thus, the sign
of the interaction contribution to the chiral condensate is governed
by a competition between these two contributions.

Our results agree with those of Kaiser and Weise~\cite{Kaisernm}, who
calculated the interaction corrections to the chiral condensate in
chiral perturbation theory with explicit $\Delta$'s including one- and
two-pion exchange contributions up to three-loop order in the energy
density. This leads to a $\approx$$\, 5 \%$ contribution at $n_0$. Extrapolating their results to higher densities, chiral symmetry
restoration is found at $\approx 3 n_0$. The
chiral condensate in neutron matter was also calculated to NLO by
Lacour {\it et al.}~\cite{Lacourcond} using in-medium chiral
perturbation theory~\cite{Lacour}. They found only small interaction
corrections, which reduce the chiral condensate at this level. Lacour
{\it et al.} also calculated the chiral condensate for the $u$ and $d$
quarks separately and showed that the condensate for the $u$ quarks is
larger than for the $d$ quarks by $\approx$$\, 7 \%$ at $n=0.2 \, {\rm
fm}^{-3}$.  This difference is however smaller than the uncertainty
from the $\sigma_{\pi N}$ term.

Finally, Kaiser \textit{et al.}~\cite{Kaiser} also computed the
interaction corrections to the chiral condensate in symmetric nuclear
matter in the same scheme. Both in neutron and symmetric matter, they
find a strong enhancement of the condensate owing to correlation
diagrams involving the excitation of the $\Delta$. In our work, the
corresponding contribution is included mainly through 3N interactions.
On a qualitative level, our results agree with those of
Refs.~\cite{Lutz,Kaiser}. However, the various interaction
contributions and the magnitude of the enhancement seem rather
different. This may be due to differences in the calculational
schemes, but also due to differences in the system considered (neutron
matter versus symmetric nuclear matter). In the future, we will study
symmetric matter, so that a direct comparison can be made.

In summary, we find that nuclear interactions impede the restoration
of chiral symmetry in neutron matter at zero temperature. The net
effect of interactions remains below $10 \%$ for $n \lesssim 0.2
\fmiq$, but grows with increasing density. The dominant source of
uncertainty is the $\sigma_{\pi N}$ term. We conclude that for moderate
densities, say $n \lesssim 0.3\fmiq$, a chiral phase transition in
neutron-rich matter therefore seems unlikely, although we cannot exclude
a strong first-order transition. For the densities considered here, we 
find a good convergence of the chiral condensate from N$^2$LO to N$^3$LO
in chiral EFT. Clearly it would be very interesting to calculate the
chiral condensate also for higher densities. While a systematic
calculation in chiral EFT is difficult at densities much higher than
$n= 0.2 \fmiq$, astrophysical observations shed light on matter at high
densities (see Refs.~\cite{Hebeler,Lattimer,longns,Weise} for the equation
of state).

\begin{acknowledgments}

We thank E.\ Epelbaum and W.\ Weise for discussions. This work was
supported by the Helmholtz Alliance Program of the Helmholtz
Association, contract HA216/EMMI ``Extremes of Density and
Temperature: Cosmic Matter in the Laboratory'', by the ERC Grant
No.~307986 STRONGINT, the DFG through Grant SFB 634, and the NSF Grant
No.~PHY--1002478.

\end{acknowledgments}

\end{document}